\documentclass[twocolumn]{revtex4}
\begin{document}

\title{\large Noncommutative Gauge Theories: Model for Hodge theory}

 \author{ Sudhaker Upadhyay}\email{sudhakerupadhyay@gmail.com}\thanks{Currently affiliated 
 at S. N. Bose National Centre for Basic Sciences, 
Block JD, Sector III, Salt Lake,   India.  }
\author{ Bhabani Prasad Mandal}\email{bhabani.mandal@gmail.com}

\affiliation { Department of Physics, 
Banaras Hindu University, 
Varanasi-221005, INDIA. 
}

\begin{abstract}
The nilpotent BRST, anti-BRST, dual-BRST and anti-dual-BRST symmetry transformations 
are constructed in the context of  noncommutative (NC) 1-form as well as 2-form  gauge 
theories.  The corresponding Noether's  charges for these symmetries 
 on the Moyal plane are shown to satisfy the same algebra as by the de Rham 
cohomological operators of
differential geometry. The Hodge decomposition theorem on compact manifold  is also studied.
We show that noncommutative  gauge theories are field theoretic models for Hodge theory. 
\end{abstract}
\maketitle

\section{Introduction}

In the past several years the noncommutative (NC) field theories (i.e. field theories on the 
Moyal plane)
 have been extensively studied from many different aspects \cite{mm,ar,ar1,dh,sw,mc,m1}. 
The motivations for NC field theories come  from the string theory. The end
points of the open strings trapped on the D-brane in the presence of two form  background 
field turn out to be
NC \cite{mm}.
The BRST symmetry in noncommutative $U(N)$ gauge theory was recently studied
 by determining the BRST transformations of the components of gauge and related fields \cite
{soro}.
In commutative gauge theories the conserved current $j_\mu(x)$ exists according to the 
Noether's theorem.
However, in the case of noncommutative field theory, it
was shown that the divergence of the current is equal to the Moyal
($\ast $) product of  some functions \cite{msj}. 
One requires  
 the physical subspace of total Hilbert space of states contains only those states that 
 are annihilated by the nilpotent and conserved BRST charge $Q_b$ i.e.  
 $Q_b\left|phys\right> =0$ in the formulation of the commutative gauge theories \cite{sund}. 
The nilpotency  of the BRST charge ($Q^2_b = 0$)
and the physicality criteria ($Q_b\left|phys \right>= 0$) are the two essential ingredients
 of BRST quantization. 
However,  the conserved charges in the NC gauge theories exist only for the spacelike 
noncommutativity. 

In the language of differential geometry  defined on compact,
 orientable
 Riemannian manifold, the cohomological aspects of BRST charge is realized in a simple, 
elegant manner. The nilpotent BRST charge is connected with  exterior derivative 
(de Rham cohomological operator $d = dx^\mu\partial_\mu,$ where $ d^2 =
0$)\cite{egu,nis,nis1,kala,hol,um}.  The conserved charge corresponding the dual-BRST 
transformation, which is also the symmetry of the action and leaves the gauge fixing 
part of the action invariant separately, is shown to be analogue of co-exterior derivative 
( $\delta = \pm \star  d \star,$ where $ \delta^2 =
0$ and $\star$ is the Hodge duality operation)\cite{um}.  

The structure of NC local groups satisfies the no-go theorem \cite{masu}.
According to this theorem, the closure condition of the gauge algebra suggests that
i) the local NC $U(N)$ algebra only admits the irreducible $N\times N$ matrix representation,
ii) and for any gauge group consisting of several single-group factors, the matter fields can
transform under at most {\it two} NC group factors. 
 In this work we consider  pure 1-form   as well as pure  2-form gauge theories in NC spacetime
which satisfy the first axiom of no-go theorem. 
We analyse the nilpotent symmetries for the 1-form gauge theory   in $n$
spacetime dimensions   and for
the 2-form gauge theory in $(1+3)$ dimensional (4D) spacetime. Some attempts in the direction
of nilpotent BRST symmetries have been made for the physical 4D  1-form gauge theories but the symmetry transformations
turn out to be nonlocal and noncovariant \cite{mc1,mc2,vi1,vi2}.
However, recently the BRST symmetry and renormalizability of the
1-form gauge theory in $(1+1)$  dimensions (2D) as well as in  4D 
have been studied thoroughly \cite{blas, blas1}. 
In the present investigation, the symmetry transformations of the
2D 1-form  and the 4D
 2-form gauge theories    are  local and covariant.
We construct the  nilpotent BRST symmetry,  anti-BRST (where the role of ghost and 
anti-ghost fields 
are changed with some changes in coefficients), dual-BRST and anti-dual-BRST transformations 
 in this framework.
 The generators of all these continuous 
symmetry transformation
are shown to obey the algebra of de Rham cohomological operators of differential geometry.
Hodge decomposition theorem in quantum Hilbert space of states is also discussed.
The BRST and anti-dual-BRST charges are mapped with exterior derivative of differential 
geometry.
On the other hand  the anti-BRST and dual-BRST charges are shown to be analogue of co-exterior 
derivative.
Further, we show that the NC 1-form as well as 2-form gauge theories are field theoretic 
models for 
Hodge theory. 

The paper is organized as follows. 
In sections II and III, we describe  the $U(N)$ 1-form and 2-form gauge theories in NC 
spacetime
respectively with their symmetry transformations. In Sec. IV, we discuss the Hodge-de Rham 
decomposition theorem 
for differential geometry. The geometrical aspects of conserved charges of
NC theories are described in Sec. V. The last section is reserved for  concluding remarks.

\section{Noncommutative 1-form gauge theory} 
\subsection{BRST and anti-BRST symmetries} 
We start with the BRST invariant action for  $n$ dimensional
1-form gauge theory (in manifestly covariant gauge) in NC spacetime as
\begin{equation}
S_B =\int d^n x\ {\cal L}_B,
\end{equation}
with Lagrangian density
\begin{eqnarray}
{\cal L}_B &=& Tr\left[ -\frac{1}{4} F_{\mu\nu}\ast F^{\mu\nu} +B\ast \partial\cdot A +\frac
{1}{2} B\ast B\right.\nonumber\\
&-&\left. i\partial_\mu \bar c \ast D^\mu c\right],\label{lag}
\end{eqnarray}
where $B, c,$ and $\bar c$ are the Nakanishi-Lautrup auxiliary field, ghost field and 
anti-ghost field 
respectively and $\ast$ is the Moyal star product.
Here the trace is taken over the $N\times N$ matrices. The field strength tensor ($F_{\mu\nu}$) 
and covariant derivative 
($ D_\mu$) of $c$ are defined as 
\begin{eqnarray}
 F_{\mu\nu} &=& \partial_\mu A_\nu - \partial_\nu A_\mu -[A_\mu, A_\nu]_\ast,\nonumber\\
 D_\mu c&=& \partial_\mu +g  A_\mu \ast c,
\end{eqnarray}
with the definition of star commutator 
\begin{eqnarray}
[A(x), B(x)]_\ast =A(x)\ast B(x)-B(x)\ast A(x).
\end{eqnarray}
The connection $A_\mu$ takes the values in the algebra of $U(N) $, with generators $T^a$ 
satisfying following
commutation and anticommutation relations 
\begin{eqnarray}
[T^a, T^b]&=&if^{abc}T^c,\nonumber\\
\{ T^a, T^b\} &=&d^{abc}T^c,
\end{eqnarray}
where $f^{abc}$ and $d^{abc}$ are  totally antisymmetric and symmetric in nature respectively.
The gauge field ($A_\mu $), ghost field ($c$), anti-ghost field ($\bar c$) and 
Nakanishi-Lautrup field 
($B$) are described in the forms of components fields as
\begin{eqnarray}
A_\mu (x) &=&    A_{\mu}^a(x) T^{a},\nonumber\\ 
c(x) &=&  c^a(x) T^{a},\nonumber\\   
\bar c(x) &=&   \bar c^a(x) T^{a},\nonumber\\
B(x) &=&    B^a(x) T^{a},
\end{eqnarray}
where group index $a$ has following values, $a\equiv 0,1,..., N^2 -1$.
The  Lagrangian density ${\cal L}_B $ remains invariant under following off-shell nilpotent
 BRST transformation ($s_b$):
\begin{eqnarray}
s_b A_\mu &=& D_\mu c,\ \ s_b c = -\frac{1}{2} g (c\ast c),\ \ s_b \bar c= i B,\nonumber\\ 
s_b B&=&0.
\end{eqnarray}
The conserved current $j_\mu $ in NC field theory is calculated by 
considering the divergence of the current is equal to the Moyal product of the certain 
functions as \cite{mj}
\begin{eqnarray}
\partial^\mu j_\mu = [f(x), g(x)]_\ast,
\end{eqnarray}
where the functions $f(x)$ and $g(x)$ are, specific to the symmetry and the star commutator, 
defined as
\begin{eqnarray}
[f(x), g(x)]_\ast &=& \left(e^{\frac{i}{2} \theta^{\mu\nu}\partial_\mu^1\partial_\nu^2}
-e^{-\frac{i}{2} \theta^{\mu\nu}\partial_\mu^1\partial_\nu^2}\right)\times \nonumber\\
&& f(x_1)g(x_2)|_{x_1 =x_2=x}.
\end{eqnarray}
The nilpotent conserved charge for the above BRST 
 transformation in NC spacetime is calculated as 
 \begin{eqnarray}
Q_{b} &=&\int d^{n-1}x \ j_0 =
 \int d^{n-1}x\left[ B\ast D_0 c\right.\nonumber\\
 & -&\left.\dot B\ast c+\frac{1}{2}ig\dot
{\bar c}\ast c\ast c\right].
\end{eqnarray}
Here we note that the  BRST transformation leads to a conserved charge only 
for the spacelike 
noncommutativity i.e. $\theta^{0i}=0$ and   our whole analysis is therefore 
restricted to this case. 

To calculate the absolutely anticommuting BRST and anti-BRST transformation, 
one needs to introduce another 
Nakanishi-Lautrup type auxiliary field ($\bar B$) in the Lagrangian density 
given in Eq. (\ref{lag}) as 
\begin{eqnarray} 
{\cal L}_{\bar B} &=& Tr\left[ -\frac{1}{4} F_{\mu\nu}\ast F^{\mu\nu} +B\ast \partial\cdot A 
+\frac{1}{2}
(B\ast B\right.\nonumber\\
&+ &\left.\bar B\ast \bar B) 
- i\partial_\mu \bar c \ast D^\mu c\right],\\
&=& Tr\left[ -\frac{1}{4} F_{\mu\nu}\ast F^{\mu\nu} +\bar B\ast \partial\cdot A +\frac{1}{2}
(B\ast B \right.\nonumber\\
&+&\left. \bar B\ast \bar B)
-i\partial_\mu \bar c \ast D^\mu c\right].\label{lag2}
\end{eqnarray} 
The BRST and anti-BRST ($s_{ab}$) symmetry transformations under which above Lagrangian 
density remains invariant
 are  
\begin{eqnarray}
s_b A_\mu &= &D_\mu c,\  s_b c = -\frac{1}{2} g  c\ast c,\ \
  s_b \bar c= i B,\nonumber\\ 
s_b B&=&0,\  s_b \bar B = g\bar B\ast c,\nonumber\\
s_{ab} A_\mu &= & D_\mu \bar c,\ s_{ab} \bar c =-\frac{1}{2} g\bar c\ast \bar c,\  
s_{ab} \bar B=0,\nonumber\\
  s_{ab} c &=&i \bar B,\ \
 s_{ab} B=  g B\ast \bar c.\label{anti}
\end{eqnarray}
where the auxiliary fields $B$ and $\bar B$ are restricted to satisfy the following 
Curci-Ferrari (CF) 
type condition \cite{cf,bt}
\begin{equation}
B+\bar B =ig (c\ast \bar c).\label{cf}
\end{equation}
The nilpotent charge corresponding to 
the anti-BRST transformation $(s_{ab})$, using Noether's theorem, is calculated as 
\begin{eqnarray}
Q_{ab} =\int d^{n-1} x\left[\dot{\bar B}\ast\bar c -\bar B\ast D_0 \bar c -\frac{1}{2}i g \dot c
\ast \bar c\ast \bar c
\right].
\end{eqnarray}

\subsection{Dual-BRST and anti-dual-BRST symmetries}
In this subsection we develop two more nilpotent symmetry transformations known as dual-BRST 
and anti-dual-BRST transformations which leave the
gauge-fixing part of the 2D Lagrangian density  in Eq. (\ref{lag}) invariant
separately.
We linearize  the kinetic part of the  Lagrangian density by introducing
 an extra auxiliary field ${\cal B}$. The linearized Lagrangian density can then be written as
\begin{eqnarray}
{\cal L}_{\cal B}& =&Tr\left[{\cal B}\ast E -\frac{1}{2} {\cal B}\ast {\cal B} +B
\ast\partial\cdot A 
+\frac{1}{2} B\ast B\right.\nonumber\\
&-&\left. i\partial_\mu \bar c\ast D^\mu c\right],
\end{eqnarray}
where $E$ is the electric field.

The off-shell nilpotent dual-BRST transformation for above Lagrangian density is given by
\begin{eqnarray}
s_d A_\mu &=&-  \epsilon_{\mu\nu} \partial^\nu \bar c,\
   s_d c =-i  {\cal B},\
   s_d \bar c=0,\nonumber\\
s_d {\cal B}&=&0,\ s_d B=0,
\end{eqnarray}
where $ \epsilon_{\mu\nu}$ is a Levi-Civita tensor of rank-2.
The conserved charge for above dual-BRST  is then calculated using Noether's theorem, as
\begin{eqnarray}
Q_d =\int dx \left[ {\cal B}\ast \dot{\bar c}-D_0{\cal B}\ast \bar c -ig\bar c
\ast\partial_1\bar c\ast
c\right].\end{eqnarray}
Here we note that the sapcelike noncommutativity in the present 2D 
reflects the Moyal star products  in the ordinary product and
therefore theory turnout to be in commutative world. 

To write the absolutely anticommuting dual and anti-dual-BRST
transformation we introduce one more auxiliary field $\bar B$.
Then the Lagrangian density given in Eq. (\ref{lag2}) is written as
\begin{eqnarray}
{\cal L}_{\bar{\cal B}} &=&Tr\left[{\cal B}\ast E -\frac{1}{2} {\cal B}\ast {\cal B} -\bar B
\ast\partial\cdot A 
\right.\nonumber\\
&+&\left.\frac{1}{2} (B\ast B +\bar B\ast\bar B)
-iD_\mu \bar c\ast \partial^\mu c\right].
\end{eqnarray}
The off-shell nilpotent anti-dual-BRST  transformation ($s_{ad}$), under which  the  
Lagrangian density ${\cal 
L}_{\bar{\cal B}}$ remains invariant,  is given by
\begin{eqnarray} 
s_{ad} A_\mu &=&-  \epsilon_{\mu\nu}\partial^\nu c,\ \ s_{ad}c =0,\ \ s_{ad}\bar c =i 
{\cal B},\nonumber\\ s_{ad} {\cal B} &=&0,\ \
 s_{ad}\bar B=0,\ \ s_{ad}\bar B=0.
\end{eqnarray}
Using Noether's  theorem, the nilpotent and conserved charge for anti-dual-BRST 
transformation is calculated as
\begin{equation}
Q_{ad}=\int dx\left[{\cal B}\ast\dot {\bar c}-D_0{\cal B}\ast \bar c-ig\bar c
\ast\partial_1\bar c\ast c\right].
\end{equation}
In the above expression of charge the Moyal product behaves as a ordinary
product. 
\subsection{Bosonic symmetry transformation} 
We construct the bosonic symmetry transformations ($s_\omega$ and  $s_{\bar\omega}$)
 for the noncommutative 
1-form gauge theory.
The  BRST ($s_{b}$), anti-BRST ($s_{ab}$), dual-BRST ($s_{d}$) and anti-dual-BRST ($s_{ad}$) 
symmetry operators which are constructed in the previous subsections satisfy 
the following algebra 
\begin{eqnarray}
\{ s_d, s_{ad}\} &=&0,\ \ \{ s_b, s_{ab}\} =0,\nonumber\\
\{ s_b, s_{ad}\} &=& 0,\ \
  \{ s_d, s_{ab}\}  =0,\nonumber\\
\{ s_b, s_d\} &\equiv &  s_\omega,\ \{ s_{ab}, s_{ad}\} \equiv s_{\bar \omega}.\label{bos}
\end{eqnarray}
The last two anticommutators  define the bosonic symmetry of the system.
Under this bosonic symmetry transformation $s_w$ the field variables transform  as
  \begin{eqnarray}
 s_\omega \bar c&=&0,\ s_\omega c=0,\ \ s_\omega B=0,\ \ s_\omega{\cal B}=0,\nonumber\\
 s_\omega A_\mu &=& -i [D_\mu {\cal B} +\epsilon_{\mu\nu}\partial^\nu B\nonumber\\
&-& ig\epsilon_{\mu\nu}
 \partial^\nu\bar c\ast c].
 \end{eqnarray}
 The conserved charge for the above bosonic symmetry transformation is calculated as
 \begin{eqnarray}
 Q_\omega &=&\int dx \left[ {\cal B}\ast\dot B -B\ast D_0{\cal B} -ig({\cal B}\ast \dot{\bar 
c}\right.\nonumber\\
 &-&\left.\partial_1\bar c\ast B)\ast c
 \right].
 \end{eqnarray}
On the other hand the bosonic symmetry transformation $s_{\bar w}$  for the field variables  
is given by
 \begin{eqnarray}
 s_{\bar \omega} \bar c&=&0,\ \ s_{\bar \omega} c=0,\ \ s_{\bar \omega} B=0,\ \ s_{\bar 
\omega}{\cal B}=0,\nonumber\\
 s_{\bar \omega} A_\mu &=& -i [-D_\mu {\cal B} +\epsilon_{\mu\nu}\partial^\nu \bar B
\nonumber\\
&-& ig\epsilon_{\mu\nu}
 \partial^\nu c\ast \bar c]
 \end{eqnarray}
Using the Noether's theorem we calculate the generator of this symmetry 
transformation 
$s_{\bar \omega}$ as 
\begin{eqnarray}
 Q_{\bar \omega} &=&\int dx \left[ {\cal B}\ast\dot {\bar B} -\bar B\ast D_0{\cal B} +ig({
\cal B}\ast \dot{  c}
\right.\nonumber\\
& -&\left. \bar B
\ast \partial_1 c)\ast \bar c
 \right].
\end{eqnarray}
 Here we note that both bosonic transformations ($s_{\bar \omega}$ and $s_{ \omega}$ ) are 
not independent 
 as their conserved charges are equivalent on the CF type restricted surface (\ref{cf}). 
 
\subsection{Ghost symmetry }
The Lagrangian density has yet another symmetry namely ghost scaling symmetry. 
The Lagrangian density as well as the ghost part of it remain invariant under the following scale transformation  for the ghost fields 
\begin{eqnarray} 
 c\longrightarrow e^{-\tau} c,\ \
\bar c \longrightarrow  e^{\tau} \bar c, 
\end{eqnarray}
where $\tau$ is a real scale parameter.
The ghost number of the ghost field ($c$) and anti-ghost field ($\bar c$)  are 1 and -1 
respectively.
Rest of the fields in the action,
whose ghost number is zero, do not change,
\begin{eqnarray} 
A_\mu &\longrightarrow &A_\mu,\ B\longrightarrow B,\ \bar B\longrightarrow \bar B. 
\end{eqnarray}
The above scale transformation leads to the following conserved ghost charge
\begin{eqnarray} 
Q_g=-i\int d x \left[\dot{\bar B}\ast \bar c -\bar B \ast D_0\bar C -\frac{1}{2} ig \dot c
\ast \bar c\ast \bar c
\right].
\end{eqnarray}
All these charges constructed in this section will be shown to satisfy the algebra satisfied by
the de Rham cohomological operators in section IV. 
\section{Noncommutative free 4D Abelian 2-form gauge theory}
The purpose of this section is to extend the results of the previous section in the case of
2-form gauge theory. 
In this section, we discuss the absolutely anticommuting BRST and anti-BRST transformations
of Abelian 2-form gauge in noncommutative plane. Dual BRST, anti-dual BRST, bosonic and ghost 
symmetry transformations 
for such theory are also
constructed.
\subsection{BRST and anti-BRST symmetry transformations}
The coupled Lagrangian densities for 2-form gauge theory in 4D \cite{manda}, which remains 
unchanged under nilpotent and absolutely
anticommuting   BRST and anti-BRST symmetry transformations, in noncommutative plane are 
given by
\begin{eqnarray}
{\cal L}_{(\beta, {\cal B})}&=& Tr\left[\frac{1}{2}\partial_\mu\varphi_2\ast 
\partial^\mu\varphi_2 -\frac{1}{2}{\cal B}^
\mu\ast 
\varepsilon_{\mu\nu\eta\kappa}\partial^\nu B^{\eta\kappa} 
\right.\nonumber\\
&-&\left.\frac{1}{2}({\cal B}_\mu\ast{\cal B}^\mu +\bar{\cal B}_\mu\ast
\bar{\cal B}^\mu)- \beta^\mu\ast \partial^\nu B_{\nu\mu}\right.\nonumber\\
&+&\left.\frac{1}{2}({ \beta_\mu}\ast{  \beta^\mu} +\bar{ \beta}_\mu\ast
\bar{ \beta}^\mu) -\frac{1}{2}\partial_\mu\varphi_1 \ast 
\partial^\mu\varphi_1\right.\nonumber\\
 &+&\left. \partial_\mu\bar \sigma 
\ast \partial^\mu\sigma
+(\partial_\mu\bar {\rho}_\nu -\partial_\nu\bar {\rho}_\mu)\ast\partial^\mu \rho^\nu 
\right.\nonumber\\
&+&\left.(\partial_\mu \rho^\mu -\chi)\ast 
\bar\chi +(\partial_\mu\bar {\rho}^\mu +\bar\chi)\ast \chi\right.\nonumber\\
 &+&\left. L^\mu\ast (\beta_\mu -\bar {\beta}_\mu -\partial_\mu \varphi_1) 
\right.\nonumber\\
&+&\left. M^\mu \ast({\cal B}_\mu -\bar {\cal B}_\mu -\partial_\mu \varphi_2 )\right],\label
{lag21} 
\end{eqnarray}
\begin{eqnarray}
{\cal L}_{(\bar \beta, \bar{\cal B})}&=& Tr\left[
 \frac{1}{2}\partial_\mu\varphi_2\ast \partial^\mu\varphi_2 -\frac{1}{2}\bar{\cal B}^\mu\ast 
\varepsilon_{\mu\nu\eta\kappa}\partial^\nu B^{\eta\kappa} \right.\nonumber\\
&-&\left.\frac{1}{2}({\cal B}_\mu\ast{\cal B}^\mu +\bar{\cal B}_\mu
\ast\bar{\cal B}^\mu)- \bar \beta^\mu\ast \partial^\nu B_{\nu\mu}\right.\nonumber\\
&+&\left.\frac{1}{2}({  \beta}_\mu\ast{ \beta}^\mu +\bar{ \beta}_\mu\ast
\bar{ \beta}^\mu)-\frac{1}{2}\partial_\mu\varphi_1\ast \partial^\mu\varphi_1 
\right.\nonumber\\
&+&\left.\partial_\mu\bar \sigma \ast  \partial^\mu\sigma
+(\partial_\mu\bar {\rho}_\nu -\partial_\nu\bar {\rho}_\mu)\ast\partial^\mu \rho^\nu 
\right.\nonumber\\
&+&\left.(\partial_\mu \rho^\mu -\chi)\ast 
\bar\chi +(\partial_\mu\bar {\rho}^\mu +\bar\chi)\ast \chi \right.\nonumber\\
&+&\left. L^\mu\ast (\beta_\mu -\bar {\beta}_\mu -\partial_\mu 
\varphi_1) \right. \nonumber\\
&+& \left. M^\mu \ast({\cal B}_\mu -\bar {\cal B}_\mu -\partial_\mu \varphi_2 )\right],\label
{lag22}
\end{eqnarray}
where the Lorentz vectors $L_\mu$ and $M_\mu$ are the Lagrange multiplier fields and ${\cal 
B}_\mu, \bar {\cal B}_\mu, 
\beta_\mu, \bar\beta_\mu$  are the Nakanishi-Lautrup type auxiliary vector fields.
The fields $\rho_{\mu}$ and $\bar\rho_{\mu}$ are anticommuting vector fields, fields $\chi$ 
and 
$\bar\chi$ are anticommuting scalar fields 
and fields $\sigma, \varphi_1,$ and $ \tilde\sigma$ are commuting scalar fields.
These fields are described in the component form as
\begin{eqnarray}B_{\mu\nu}(x) &=&  B_{\mu\nu}^a(x) T^{a},\ \  
\rho_\mu (x) =  \rho_\mu^a (x) T^{a},\nonumber\\   
\bar\rho_\mu(x) &=& \bar\rho_\mu(x)^a(x) T^{a},\ \
\sigma(x) =  \sigma^a(x) T^{a},\nonumber\\
\bar\sigma(x) &=&   \bar\sigma^a(x) T^{a},\ \ 
\chi(x) = \chi^a(x) T^{a},\nonumber\\
\bar\chi(x) &=&   \bar\chi^a(x) T^{a},\ \ 
\beta_\mu(x) =  \beta_\mu^a(x) T^{a},\nonumber\\
\bar\beta_\mu(x) &=&   \bar\beta_\mu^a(x) T^{a},\ \   
{\cal B}_\mu(x) = {\cal B}_\mu^a(x) T^{a},\nonumber\\
\bar{\cal B}_\mu(x) &=&  \bar{\cal B}_\mu^a(x) T^{a},\ \ \varphi_1(x) =  \varphi_1^a(x) 
T^{a},\nonumber\\
\varphi_2(x) &=& \varphi_2^a(x) T^{a},\ \ 
L_\mu(x) =L_\mu^a(x) T^{a},\nonumber\\
M_\mu(x) &=& M_\mu^a(x) T^{a}.
\end{eqnarray}
The above two coupled Lagrangian densities in Eqs. (\ref{lag21}) and (\ref{lag22}) are equivalent on the following CF type 
restricted surface
\begin{eqnarray}
{\cal B}_\mu -\bar {\cal B}_\mu -\partial_\mu \varphi_2 =0, \ \beta_\mu -\bar {\beta}_\mu -
\partial_\mu \varphi_1.
\end{eqnarray}
The above Lagrangian densities for 2-form gauge theory are invariant under the following off-
shell nilpotent   and absolutely anticommuting   BRST and anti-BRST transformations  as
\begin{eqnarray}
s_b B_{\mu\nu} &=& -(\partial_\mu \rho_\nu -\partial_\nu \rho_\mu ), \ s_b \rho_\mu =-
\partial_\mu \sigma,\nonumber\\
  s_b \bar \rho_\mu
&=&-\beta_\mu,\
 s_b \bar \sigma =-\bar\chi, \  s_b \bar \beta_\mu =-\partial_\mu\chi,\nonumber\\  s_b 
\varphi_1 &=&\chi,\
s_b L_\mu =-\partial_\mu\chi,\ 
 s_b \alpha  =0,\nonumber\\ (\alpha &\equiv & \bar\chi, \chi, \sigma, \varphi_2, \beta_\mu, {
\cal B}_
\mu, 
\bar {\cal B}_\mu, M_\mu),
\end{eqnarray}
\begin{eqnarray}
s_{ab} B_{\mu\nu} &=& -(\partial_\mu \bar \rho_\nu -\partial_\nu \bar \rho_\mu ),\ s_{ab} 
\bar \rho_\mu =-\partial_\mu 
\bar \sigma,
\nonumber\\
  s_{ab} \rho_\mu &=&\bar \beta_\mu,\ \
 s_{ab} \sigma =-\chi, \   s_{ab}  \beta_\mu =\partial_\mu\bar\chi,\ 
\nonumber\\
s_{ab}\varphi_1 &=&\bar\chi,\ 
s_{ab} L_\mu =-\partial_\mu\bar\chi,\ \
 s_{ab} \gamma  =0,\nonumber\\ (\gamma &\equiv &\bar\chi, \chi, \bar\sigma, \varphi_2, {\cal 
B}_\mu, 
\bar {\cal B}_\mu, 
\bar {  \beta}_\mu, M_\mu), 
\end{eqnarray}
The Noether's charges for above BRST and anti-BRST symmetries are calculated as 
\begin{eqnarray}
Q_b &=& \int d^3x \left[(\partial_0\bar\rho_\nu - \partial_\nu\bar \rho_0 )\ast  
\partial^\nu\sigma \right.\nonumber\\
&-&\left.\epsilon^{0
\nu\eta\kappa}(\partial_\nu\rho_\eta)\ast {\cal B}_\kappa 
-\bar\chi\ast \partial_0\sigma\right.\nonumber\\
 &-&\left.(\partial_0\rho_\nu -\partial_\nu 
\rho_0 )\ast \beta^\nu -\chi\ast \partial_0\varphi_1\right.\nonumber\\
&-&\left.\chi \ast L_0 \right],
\end{eqnarray}
\begin{eqnarray}
Q_{ab} &=& \int d^3x \left[-(\partial_0\rho_\nu - \partial_\nu \rho_0 )\ast 
\partial^\nu\bar\sigma \right.\nonumber\\
&-&\left.\epsilon^{0
\nu\eta\kappa}(\partial_\nu\bar\rho_\eta) \ast \bar{\cal B}_\kappa 
-\chi\ast \partial_0\bar\sigma\right.\nonumber\\ &-&\left.
(\partial_0\bar\rho_\nu -
\partial_\nu\bar\rho_0 )\ast \bar\beta^\nu 
-\bar\chi\ast \partial_0\varphi_1\right.\nonumber\\
 & -&\left.\bar\chi\ast  L_0 \right].
\end{eqnarray}
These charges will be used in section IV.
\subsection{Dual and anti-dual-BRST symmetries}
The
dual and anti-dual-BRST transformations  are also the symmetries of the
effective action for 2-form gauge theory. Further these transformations leave 
the gauge fixing term invariant independently.   
The nilpotent and absolutely anticommuting dual-BRST and anti-dual-BRST transformations, 
which leave this 
noncommutative 
2-form Lagrangian density invariant, are calculated as follows,
\begin{eqnarray} 
&&s_dB_{\mu\nu}= -\epsilon_{\mu\nu\eta\kappa}\partial^\eta\bar \rho^\kappa,\ s_d \bar 
\rho_\mu =-\partial_\mu\bar \sigma,\nonumber\\
&&s_d \rho_\mu = -{\cal B}_\mu,
\ s_d \varphi_2 = -\bar\chi,\  s_d\sigma =-\chi, \nonumber\\
&&  s_d\bar{\cal B}_\mu =\partial_\mu\bar\chi,\
s_d M_\mu =-\partial_\mu \bar\chi,\ \
s_d \varrho = 0,\nonumber\\
&& (\varrho\equiv \bar\chi, \chi, \bar\sigma, \varphi_1, {\cal B}_\mu,
 \beta_\mu, \bar { \beta}_\mu, L_\mu),
\end{eqnarray}
\begin{eqnarray}
&&s_{ad}= -\epsilon_{\mu\nu\eta\kappa}\partial^\eta  \rho^\kappa,\ s_{ad}   \rho_\mu = 
\partial_\mu  
\sigma,\nonumber\\
&&s_{ad} \bar \rho_\mu =  \bar{\cal B}_\mu,\ \
s_{ad} \varphi_2 = -\chi,\  s_{ad}\sigma = \bar\chi, \nonumber\\
&&  s_{ad} {\cal B}_\mu =-\partial_\mu\lambda,\
s_{ad} M_\mu =-\partial_\mu \chi,\ s_{ad} \varphi =0,\nonumber\\
&& (\varphi\equiv \bar\chi, \chi, \sigma, \varphi_1, \bar{\cal B}_
\mu, {\beta}_\mu, \bar {\beta}_\mu, L_\mu).
\end{eqnarray}
The Noether's charges for above dual BRST and anti-dual-BRST symmetries are calculated as 
\begin{eqnarray}
Q_d &=& \int d^3x \left[(\partial_0\bar\rho_\nu - \partial_\nu\bar \rho_0 )\ast {\cal B}^\nu 
\right.\nonumber\\
&-&\left.\epsilon^{0
\nu\eta\kappa}\beta_\nu \ast (\partial_\eta\bar\rho_\kappa) 
-\bar\chi\ast \partial_0\varphi_2 \right.\nonumber\\
&-&\left.(\partial_0\rho_\nu -\partial_\nu \rho_0 )\ast \partial^\nu\bar\sigma 
\right.\nonumber\\
& -&\left.\chi\ast \partial_0\bar\sigma +\bar\chi\ast  M_0 \right],
\end{eqnarray}
\begin{eqnarray}
Q_{ad} &=& \int d^3x \left[(\partial_0\rho_\nu - \partial_\nu \rho_0 )\ast \bar{\cal B}^\nu
\right.\nonumber\\
&-&\left.\epsilon^{0
\nu\eta\kappa}\bar\beta_\nu\ast (\partial_\eta\rho_\kappa) 
-\chi\ast \partial_0\varphi_2\right.\nonumber\\
 &-&\left.(\partial_0\bar\rho_\nu -\partial_\nu \bar\rho_0 )\ast \partial^\nu\sigma 
\right.\nonumber\\
& +&\left.\bar\chi\ast \partial_0\sigma+\chi \ast M_0 \right].
\end{eqnarray}
\subsection{Bososnic symmetry transformation}
Now we construct the bosonic symmetry transformations out of these nilpotent BRST
symmetries for this theory. The BRST ($s_b$), anti-BRST (
$s_{ab}$), dual-BRST ($s_d$) and anti-dual-BRST ($s_{ad}$) 
symmetry operators satisfy the following algebra
\begin{eqnarray}
\{s_b, s_{ab}\} &=& 0,\ \ \{s_b, s_{ad}\} =  0,\nonumber\\
\{s_d, s_{ab}\} &=&  0,\ \  \{s_d, s_{ad}\} =0,\nonumber\\
\{s_b, s_{ d}\} &=& s_\omega, \ \{s_{ab}, s_{a d}\}  =   s_{\bar\omega}.
\end{eqnarray}
The last two anticommutators define the bosonic transformations under which the fields 
transform as
\begin{eqnarray}
&&s_\omega B_{\mu\nu}=\partial_\mu {\cal B}_\nu -\partial_\nu {\cal B}_\mu +\varepsilon_{
\mu\nu\eta\kappa}\partial^\eta
\beta^\kappa,\nonumber\\
&&s_\omega \rho_\mu  =  \partial_\mu \chi,\
 s_\omega \bar \rho_\mu =\partial_\mu\bar\chi,\ 
 s_\omega \varsigma  =0,\nonumber\\
  &&(\varsigma \equiv  \bar\chi, \chi, \varphi_1, \varphi_2, \sigma, \bar \sigma, 
 \beta_\mu, \bar\beta_\mu, {\cal B}_\mu, \bar {\cal B}_\mu,\nonumber\\
&& L_\mu, M_\mu).
 \end{eqnarray}
 \begin{eqnarray}
&&s_{\bar \omega} B_{\mu\nu}=-(\partial_\mu \bar{\cal B}_\nu -\partial_\nu \bar{\cal B}_\mu +
\varepsilon_{\mu\nu\eta\kappa
}\partial^\eta
\bar \beta^\kappa),\nonumber\\
&&s_{\bar \omega} \rho_\mu  =  -\partial_\mu \chi,\ 
 s_{\bar\omega} \bar \rho_\mu =-\partial_\mu\bar\chi,\ \
 s_{\bar \omega} \varpi  =0,\nonumber\\
 &&(\varpi   \equiv \bar\chi, \chi, \varphi_1, \varphi_2, \sigma, 
\bar \sigma, 
 \beta_\mu, \bar \beta_\mu, {\cal B}_\mu, \bar {\cal B}_\mu,\nonumber\\
&& L_\mu, M_\mu).
 \end{eqnarray}
 The nilpotent and conserved charge for $s_\omega$, using Noether's theorem, is calculated as
\begin{eqnarray} 
Q_\omega &=& \int d^3x \left[\varepsilon_{0\nu\eta\kappa}\{(\partial^\nu{\cal B}^\eta)\ast {
\cal B}^\kappa + 
(\partial^\nu\beta^\eta) \ast \beta^\kappa \}
\right.\nonumber\\
&+&\left. \partial_\nu (\beta^0\ast {\cal B}^\nu -\beta^\nu\ast {\cal B}^0 )\right.\nonumber\\
&+&\left.
(\partial_0\bar\rho_\nu -\partial_\nu\bar\rho_0 )\ast \partial^\nu\chi\right.\nonumber\\ &-&
\left.(\partial_0 
\rho_\nu -\partial_\nu\rho_0 )\ast 
\partial^\nu\bar\chi \right].
\end{eqnarray}
  
\subsection{Ghost and discrete symmetry}
Now, we would like to mention  yet another symmetry of this system, namely, the ghost symmetry.
We introduce a scale transformation of the ghost field, under 
which the effective action for NC 2-form gauge theory is invariant, as 
\begin{eqnarray}
&&s_g\sigma = 2\tau  \sigma,\ s_g \bar \sigma=-2\tau\bar \sigma,\ s_g c_\mu =
\tau\rho_\mu,\nonumber\\
&&s_g \bar \rho_\mu = -\tau \bar \rho_\mu,\ s_g \bar\chi =\tau\bar\chi,\ s_g \chi =\tau\chi,\ 
s_g \varepsilon  =0,\nonumber\\
 &&\varepsilon   \equiv  \{B_{\mu\nu}, \varphi_1, \varphi_2, \beta_\mu, \bar \beta_\mu, {\cal 
B}_\mu, \bar {\cal B}_
\mu, L_\mu, M_\mu\},
\end{eqnarray}
where $\tau$ is an arbitrary scale parameter.

The conserved charge for the above symmetry transformations is 
\begin{eqnarray}
Q_g &=& \int d^3x \left[2\sigma\ast \partial^0\bar\sigma -2\bar\sigma\ast \partial^0\sigma 
\right.\nonumber\\
&+&\left. (\partial^0\rho^\nu -
\partial^\nu\rho^0 )\ast \bar\rho_\nu +(\partial^0\bar\rho^\nu -
\partial^\nu\bar\rho^0 )\ast \rho_\nu \right.\nonumber\\
&+&\left.\rho^0\ast \bar\chi -\bar\rho^0\ast \chi\right].
\end{eqnarray}
Further, the Lagrangian densities given in Eqs (\ref{lag21}) and (\ref{lag22}) remain invariant under 
following discrete 
symmetry transformations
\begin{eqnarray}
B_{\mu\nu}&\longrightarrow & \mp \frac{i}{2} \varepsilon_{\mu\nu\eta\kappa} B^{\eta\kappa}, 
\ \rho_\mu\longrightarrow 
\pm i\bar \rho_\mu,\nonumber\\
 \bar \rho_\mu &\longrightarrow & \pm i \rho_\mu,\
\sigma \longrightarrow \pm i\bar \sigma,\ \bar\sigma\longrightarrow \mp i\sigma,\nonumber\\
 \varphi_1 &\longrightarrow & \pm i
\varphi_2,  \
\varphi_2\longrightarrow \mp i \varphi_1,\ 
\bar\chi \longrightarrow\mp i\chi,\nonumber\\
 \chi &\longrightarrow &\mp i\bar\chi,\ L_\mu \longrightarrow \mp iM_\mu,\
M_\mu \longrightarrow \pm iL_\mu, \nonumber\\
\beta_\mu &\longrightarrow  &\pm i {\cal B}_\mu,\ {\cal B}_\mu\longrightarrow \mp i
\beta_\mu,\ \bar \beta_\mu 
\longrightarrow 
\pm i {\bar \beta}_\mu, \nonumber\\
\ \bar {\cal B}_\mu &\longrightarrow &\mp  i\bar {\cal B}_\mu.\label{dis}
\end{eqnarray}
The above symmetry transformations play very important role in establishing
a connection between the symmetries on the one hand and some key concepts
of the differential geometry on the other. For instance, these discrete symmetry 
transformations are the analogue of the Hodge duality operator ($\star$){\footnote{One should not 
confuse the
Moyal star product ($\ast$) with the Hodge duality operation ($\star$).}}
of differential geometry.  It is interesting to point out the following 
relations under the two successive $\star$ operations on the fields 
\begin{eqnarray}
\star  (\star  B)  =B,\ B &\equiv &\{ B_{\mu\nu}, \beta_\mu, \bar \beta_\mu, {\cal B}_\mu, \bar
 {\cal B}_\mu,\nonumber\\
 &&\varphi_1, 
\varphi_2, L_
\mu, M_\mu,
\sigma, \bar\sigma \},\nonumber\\
\star  (\star  F)=-F,\ F&\equiv & \{\rho_\mu, \bar \rho_\mu, \bar\chi, \chi\},
\end{eqnarray}
where $\star$ corresponds to the discrete symmetry transformations given in Eq. (\ref{dis}).

Thus, we note that the fermionic and bosonic fields of the theory transform 
in a different manner under the successive operations of the
discrete transformations. This important observation leads to the 
following operator relationships:
\begin{eqnarray}
s_d  = \pm \star  s_b\star ,\ s_{ad} =\pm \star  s_{ab}\star.  
\end{eqnarray}
here we note
that the above relationship is the analogue of the relationship between the
cohomological operators $\delta$ and $d$, i.e. $\delta = \pm \star  d \star$. 
\section{Hodge-de Rham decomposition theorem and differential operators}
The de Rham cohomological operators (exterior derivative $d$, co-exterior derivative $\delta$
 and Laplace-Beltrami 
operator $\Delta$) 
 of differential geometry obey the following algebra
\begin{eqnarray}
 &&d^2 = \delta^2 =0, \ \ \Delta =(d +\delta )^2 =d\delta +\delta d\equiv \{d,\delta\}
\nonumber\\
&& [\Delta, \delta ]=0, \ \ [\Delta,d ]=0, \label{alg}
\end{eqnarray}
The operators $d$ and $\delta$ are adjoints or duals of each other and  $\Delta$ 
is self-adjoint operator.
It is well-known that the exterior derivative raises the degree of a form by
one when it operates on it (i.e. $df_n\sim f_{n+1}$). On the other hand, the dual-exterior 
derivative
lowers the degree of a form by one when it operates on forms (i.e. $\delta f_n\sim f_{n-1}$). 

Let $M$ be a compact, orientable Riemannian manifold, then an inner product on the vector 
space 
$E^n (M)$ of $n$-forms on $M$ can be defined as 
\begin{equation}
(\alpha, \beta ) =\int_M \alpha\wedge \star  \beta,
\end{equation}
for $\alpha, \beta\in E^n (M)$ and $\star  $ is the Hodge duality operator \cite{mor}. 
Suppose that $\alpha$ and $\beta$ are forms of degree $n$ and $n+1$ respectively.
Then following relation for inner product will be satisfied
\begin{equation}
(d\alpha, \beta )=(\alpha, \delta\beta ).
\end{equation}
Similarly, if $\beta$ is form of degree $n-1$, then we have the relation 
$(\alpha, d\beta )=(\delta\alpha, \beta )$. Thus the necessary and 
sufficient condition for $\alpha$ 
to be closed is that it should be orthogonal to all co-exact forms of degree $n$.
The form $\omega\in E^n (M)$ is called harmonic if $\Delta \omega =0$. Now let $\beta$ be a 
$n$-form  on $M$ and if there exists another $n$-form $\alpha$ such that $\Delta\alpha=\beta$,
then for a harmonic form $\gamma\in H^n$, 
\begin{equation}
(\beta, \gamma)=(\Delta\alpha, \gamma)=(\alpha, \Delta\gamma)=0,\label{Del}
\end{equation}
where $H^n(M)$ denote the subspace of $E^n(M)$ of harmonic forms on $M$.
Therefore, if a form $\alpha$ exist with the property that $\Delta\alpha =\beta$,
then Eq. (\ref{Del}) is necessary and sufficient condition for $\beta$ to be orthogonal
 to the subspace $H^n$.
This reasoning lead to the idea that $E^n(M)$ can be partitioned in to three 
distinct subspaces $\Lambda^n_d$, $\Lambda^n_\delta$ and $H^n$ which are consistent
with exact, co-exact  and harmonic forms respectively.
The Hodge-de Rham decomposition theorem can be stated as \cite{wan}: 

 {\it A regular differential form of degree $n$ may be uniquely decomposed into a sum of
 the form
\begin{equation}
\alpha =\alpha_H +\alpha_\delta +\alpha_d,
\end{equation}
where $\alpha_H\in H^n, \alpha_\delta\in \Lambda^n_\delta$ and $\alpha_d\in \Lambda^n_d$}.

\subsection{Hodge-de Rham decomposition theorem and conserved charges}
In this subsection we study the analogy between the de Rham 
cohomological operators and the conserved charges for symmetry transformations  for noncommutative gauge theories. 
In particular we draw the similarity between the algebras obeyed by de Rham 
cohomological operators and the conserved charges.

The constructed nilpotent symmetry transformations (in earlier sections) 
for NC 2-form theory pursue the following algebra
\begin{eqnarray}
&& s_b^2  = 0,\ s_{ab}^2 =0, \ \{s_b, s_{ad}\} =0 =\{s_d, s_{ad}\},\nonumber\\
&&s_\omega = \{s_b, s_d\}\equiv -\{s_{ad}, s_{ad}\},\ [s_\omega, s_r]=0,\nonumber\\
  && [s_g, s_b ] =  s_b,\ [s_g, s_d] =-s_d,\ [s_g, s_{ad}] =s_{ad},\nonumber\\
  && [s_g, s_{ab}]  =  -s_{ab},\ \  s_r\equiv s_b, s_{ab}, s_d, s_{ad}, s_g.
  \label{fgs}
\end{eqnarray}
With the Eqs. (\ref{alg}) and (\ref{fgs}), we  draw the following two to one mappings 
\begin{eqnarray}
(s_b, s_{ad})&\longrightarrow &d,\ \ (s_d, s_{ad})\longrightarrow \delta,\nonumber\\
\{s_b, s_d\}&=& -\{s_{ab}, s_{ad}\}\longrightarrow \Delta.
\end{eqnarray}
The conserved charges of all the 
symmetry transformations
are shown to satisfy the following algebra 
\begin{eqnarray} 
&&Q_{b}^2=0,\ Q_{ab}^2=0,\ Q_{d}^2=0, \ Q_{ad}^2=0,\nonumber\\
&& \{ Q_b, Q_{ab}\} =0,\
\{ Q_d, Q_{ad}\} =0,\ \ \{ Q_b, Q_{ad}\} =0, \nonumber\\
&&\{ Q_d,Q_{ab}\}  = 0,\ 
 i[ Q_g, Q_b] = Q_b,\nonumber\\ 
 &&i[ Q_g, Q_{ad}]= Q_{ad},\ 
 i[ Q_g, Q_d] = -Q_d, \nonumber\\
&& i[ Q_g, Q_{ab}]= -Q_{ab},\ \
i[ Q_\omega, Q_r]  = 0,\nonumber\\
&&Q_r \equiv Q_b, Q_{ab}, Q_d, Q_{ad}, Q_g. \label{cgs}
\end{eqnarray} 
This algebra is reminiscent of the algebra satisfied by the de Rham cohomological operators
of differential geometry given in Eq. (\ref{alg}). Comparing (\ref{alg}) and (\ref{cgs}) we 
obtain following mappings
\begin{equation}
(Q_b, Q_{ad})\rightarrow d, \ \ (Q_d, Q_{ab})\rightarrow \delta, \ \ Q_\omega\rightarrow 
\Delta.
\end{equation}

Let $n$ be the ghost number associated with a particular state $\left|\psi\right>_n$ 
defined in the total Hilbert space of states, i.e.,
\begin{equation}
Q_g\left|\psi\right>_n = n\left|\psi\right>_n
\end{equation}
Then it is easy to verify the following relations 
\begin{eqnarray}
i Q_g Q_b\left|\psi\right>_n &=& (n+1)Q_b\left|\psi\right>_n,\nonumber\\
i Q_g Q_{ad}\left|\psi\right>_n &=& (n+1)Q_{ad}\left|\psi\right>_n,\nonumber\\
i Q_g Q_d\left|\psi\right>_n &=& (n-1)Q_d\left|\psi\right>_n,\nonumber\\
i Q_g Q_{ab}\left|\psi\right>_n &=& (n-1)Q_{ab}\left|\psi\right>_n,\nonumber\\
i Q_g Q_\omega\left|\psi\right>_n &=& nQ_\omega\left|\psi\right>_n,\label{gh}
\end{eqnarray}
which imply that the ghost numbers of the states 
$Q_b\left|\psi\right>_n$, $Q_d\left|\psi\right>_n $ and $Q_\omega\left|\psi\right>_n $
are $(n + 1), (n - 1)$ and $n $ respectively.
The states $ Q_{ab}\left|\psi\right>_n$ and $ Q_{ad}\left|\psi\right>_n $
have ghost numbers $ (n - 1)$ and $(n + 1)$ respectively. 
The properties of $d$ and $\delta$ are mimicked by sets ($Q_b,Q_{ad}$) and ($Q_d,Q_{ab}$), 
respectively. It is evident from Eq. (\ref{gh}) that  the set ($Q_b,Q_{ad}$) raises 
the ghost number of a
state by one and on the other hand the set ($Q_d, Q_{ab}$) lowers the ghost number of 
the same state
by one. These observations, keeping the analogy with the Hodge-de Rham 
decomposition theorem, enable us to express any arbitrary state 
$\left|\psi\right>_n$ in terms of the sets
 ($Q_b, Q_d, Q_\omega$) and ($Q_{ad}, Q_{ab}, Q_\omega$) as
\begin{equation}
\left|\psi\right>_n = \left|\omega\right>_n +Q_b\left|\chi\right>_{n-1}+Q_d\left|\phi\right>_
{n+1},
\end{equation}
\begin{equation}
\left|\psi\right>_n = \left|\omega\right>_n +Q_{ad}\left|\chi\right>_{n-1}+Q_{ab}
\left|\phi\right>_
{n+1},
\end{equation}
where the most symmetric state is the harmonic state $\left|w\right>_n$ which satisfies the 
following
relations,
\begin{equation}
Q_\omega\left|\omega\right>_n=0, \ \ Q_{(a)b}\left|\omega\right>_n=0, \ \ Q_{(a)d}
\left|\omega\right>_n=0,
\end{equation}
analogous to the Eq. (\ref{Del}).

\section{conclusion} 
We have considered the  NC $U(N)$ 1-form as well as $U(N)$ 2-form  
gauge theories on Moyal plane which satisfies a no-go theorem by restricting the gauge fields 
to have $N\times N$ matrix representation.
We have studied  the BRST symmetry transformation for these theories on 
Moyal plane.  Further, we have shown that the dual of BRST 
transformation, which is the
symmetry of the effective Lagrangian density and leaves the gauge-fixing part invariant 
separately,
also exists for such theories. 
Interchanging the role of ghost and antighost field with some coefficients
 anti-BRST and anti-dual-BRST symmetry
 transformations have also been constructed. 
We have noted  that all the conserved charges for such symmetry transformations  exist
only in the case of spacelike noncommutativity  ($\theta_{0i} =0$).
However, we have observed that the Moyal star product in the
expressions of  the conserved charges,
in the case of 2-dimensional NC 1-form gauge theory,
behaves like an ordinary product. 
The nilpotent BRST  symmetry transformation is turned out to be the analogue of
the exterior derivative as the kinetic term remains invariant under this. In the similar
 fashion we have shown that the dual-BRST symmetry transformation  is also
 linked with the co-exterior derivative.
 The anticommutator of either BRST and the dual-BRST symmetry generators or
 anti-BRST and anti-dual-BRST symmetry generators 
 leads to a bosonic symmetry in the theory which turns out to be
the analogue of the Laplacian operator. Further, the effective theory has a non-nilpotent
 ghost symmetry transformation which is also the symmetry of the ghost terms of the effective 
action. We have shown that the algebra satisfied by the nilpotent charges is exactly
same as the de Rham cohomological operator. These results are shown for 
for both NC 1-form and 2-form gauge theories.
 Thus, the NC
1-form as well as 2-form gauge theories  have been realized as the  field theoretic 
models for Hodge theory. It will be interesting to see that whether no-go theorem on NC spacetime puts more restrictions 
on  the matter sector of the NC gauge theories in the context of Hodge theorem. 

\begin{acknowledgments}
SU gratefully acknowledges the financial support from the Council of Scientific and 
Industrial Research
(CSIR), India, under the SRF scheme. 
\end{acknowledgments}

\end{document}